\documentclass[aps,prl,twocolumn]{revtex4}
\usepackage{graphicx,amsmath,amssymb,bm}

\begin{document}

\title{Length-scale  independent skyrmion and meron Hall angles 
	}
\author{Sandip Bera$^1$ and Sudhansu S. Mandal$^{1,2}$ }
\affiliation{$^1$Department of Physics, Indian Institute of Technology, Kharagpur, West Bengal 721302, India\\ 
$^2$Centre for Theoretical Studies, Indian Institute of Technology, Kharagpur, West Bengal 721302, India	
	}	

\begin{abstract}
Motivated by the recent observation [Zeissler {\em et al.}, Nature Comm. 11, 428 (2020)] of enigmatic radius-independent skyrmion Hall angle in chiral magnets, we derive skyrmion Hall angle based on the recent solution of skyrmions characterized by the sole length scale determined with the Dzyaloshinskii-Moriya interaction strength and applied magnetic field.
We find that the skyrmion Hall angle is independent of input current density and the length-scale which determines the radius of a skyrmion.
 This is corroborated with the single length-scale dependent skyrmion profile which is the solution of the Euler equation of polar angle representing magnetization. Although the magnitude of Hall angle may change with the change of profile (shape) of the skyrmion, it remains unchanged for a particular profile.
With the application of tunable current along mutually perpendicular directions, this property enables us to propose an experimental setup by which the transverse motion of a skyrmion can be restricted so that the skyrmion can only traverse longitudinally. We further find the length-scale and input-current density independent Hall angles for merons where their transverse motion will be opposite depending on whether the spin at their centers are up or down, in agreement with an experiment.

\end{abstract}

\maketitle

{\it Introduction}: The topologically non-trivial spin textures \cite{Nagaosa13,Fert17,Fert13} as skyrmions \cite{Muhlbauer09} and merons\cite{Phatak12} realized in chiral magnets  have opened a new avenue for fundamental physics and also a new dimension to promising routs of spintronic applications \cite{Iwasaki13,Parkin08,Du15,Beg15,Yuan16,Zhang15,Schulz2012,Jonietz1648}. While the skyrmions have topological charge \cite{Nagaosa13,Fert17,BM} $Q=1$, merons are classified \cite{BM,Hirata2019} with $Q= \frac{1}{2}$ because of which these are also known as half-skyrmions in literature. Apart from observing Bloch type skyrmions in bulk \cite{Muhlbauer09} chiral materials, interfacial inversion symmetry breaking in ferromagnetic thin films or heterostructures leading to Rashba-type Dzyaloshinskii-Moriya interaction has facilitated the observation of robust chiral spin textures such as N\'eel type skyrmions and merons \cite{Parkin08,Iwasaki13,Du15,Beg15,Yuan16,Zhang15,MoreauLuchaire2016,Soumyanarayanan2017,Schulz2012,Jonietz1648,Yu2012,Grollier03,Zhang20,Phatak12,Zeissler20,Lepadatu,Hrabec,Jiang}.

Analogous to the transverse motion of charged particles due to Lorenz force in classical Hall effect, the spin-textured quasi-particles like skyrmions and merons with topological charge have transverse motion along with usual longitudinal motion due to magnus force \cite{Landau,Gilbert,Slonczewski96}; the corresponding effect respectively known as skyrmion and meron Hall effect. Just as applied electric field is the driving force for conventional Hall effect, it is the spin-Hall current induced by the applied current \cite{Lepadatu,Zhou19,Tomasello14,Menezes1,Menezes2,Yuan19} drives the motion of a skyrmion or meron. Their motion, however, are restricted due to the dissipative force arising from intrinsic magnetic damping called Gilbert damping\cite{Landau,Gilbert,Slonczewski96}. The ratio between transverse and longitudinal velocities determines the Hall angle.

Early theoretical studies \cite{Zhou19,Wang18,Jiang17,Tomasello14,Juge19,Litzius2017,Liu19,Yuan19}
predict the skyrmion-radius dependent skyrmion Hall angle. On the contrary, Zeissler et al \cite{Zeissler20} have recently observed radius-independent  skyrmion-Hall angle for skyrmions varying diameters ranging from 35 to 825 nm.
Reichhardt and Reichhardt \cite{Reichhardt2020} have attributed this with plastic motion of skyrmions due to their strong interactions with the pinning centers. We, however, here find that the skyrmion profiles with single characteristic length scale may lead to the length-scale independent skyrmion Hall angle.

 We recently have reported an approximate closed-form solution of Euler equation for a skyrmion (meron) \cite{BM}where skyrmion-(meron-)radius is characterized by a  single length-scale derivable from the ratio of  Dzyaloshinskii-Moriya interaction strength and applied magnetic field (magnetic anisotropy). Our primary focus in this paper is to determine skyrmion (meron) Hall angle by solving Thiele equations for the skyrmion (meron) velocities with respect to these solutions. We find that the skyrmion-(meron-)Hall angle  $\theta_{\rm sk}$ depends only on the Gilbert damping constant $\alpha$, the topological charge $Q$, and a constant $S$ depending on the corresponding topological texture of the quasiparticle:
$\theta_{\rm sk} = \vert \arctan (\frac{\alpha S}{Q}) \vert$. In other words, the skyrmion (meron) Hall angle neither depends on its length scale nor on the driving current, in consistence with the experiments \cite{Zhang20,Zeissler20,Woo}.  
Although the Hall angle may increase with the increase of driving current due to the pinning effect caused by disorder, the Hall angle eventually saturates to its zero-disorder value \cite{Muller2015,Reichhardt2015,Diaz2017,Reichhardt2015b,Stosic17,Hanneken,Bhattacharya}.

One of the goals for spintronic application of skyrmions is to drive them through a narrow channel. However, due to their transverse motion arising from the skyrmion-Hall effect, they soon arrive at the boundary of the channel and their spin-textures get destroyed\cite{Sampaio}. Several experimental \cite{Zhang2016,Hirata2019} attempts and theoretical \cite{Akosa} proposals have been made for holding their longitudinal motions only. We here propose an experimental setup (Fig.~1) by applying transverse spin-polarized current in controlled manner along with the longitudinal spin-polarized current, the skyrmion Hall angle can be manipulated and zero Hall angle condition is obtained when the ratio of transverse current to the longitudinal current becomes $ \frac{\alpha S}{Q}$. The method of injecting spin-polarized current along two mutually perpendicular directions may be adopted with the technique developed earlier \cite{Tan09,Cowburn17}.

 
   
\begin{figure}
	\includegraphics[scale=0.4]{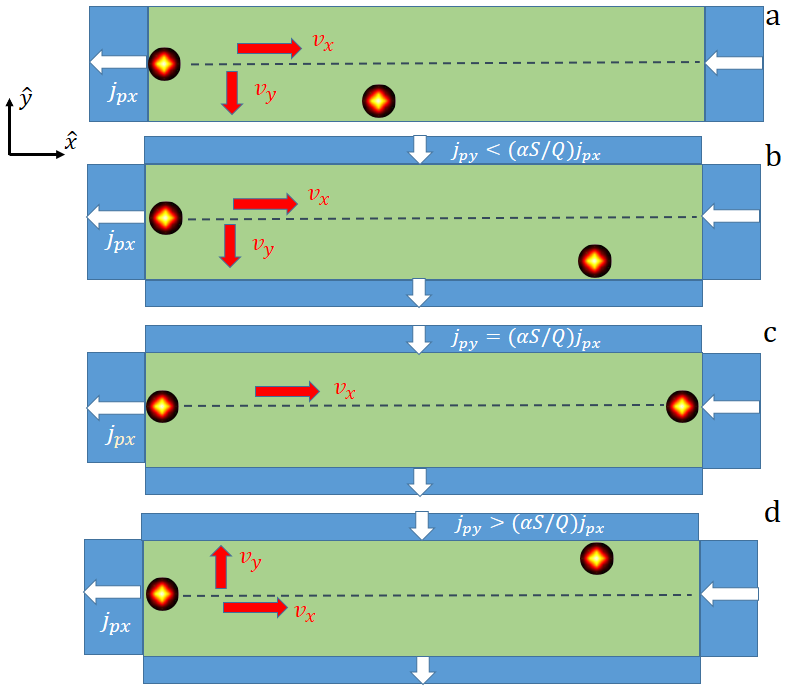}
	\caption{Schematic arrangement for the proposed experimental setup for manipulating skyrmion Hall angle. $j_{px}$ and $j_{py}$ denote applied spin-polarized current (polarization along the direction of the current) through gates along x any y directions respectively. The dark and bright spot represents the position of a skyrmion. (a) Usual $\theta_{\rm sk}$  in the absence of $j_{py}$. (b) $\theta_{\rm sk}$ decreases with the increase of $j_{py}$. (c) $\theta_{\rm sk}=0$ when $j_{py} = \frac{\alpha S}{Q}j_{px}$. (d) $\theta_{\rm sk}$ changes sign as skyrmions gets deflected in the opposite direction when $j_{py} > \frac{\alpha S}{Q}j_{px}$. }
\end{figure}

{\it Preliminaries}: The current-driven spin dynamics are generally studied by the well-known Landau-Lifshitz-Gilbert-Slonczewski (LLGS) equation \cite{Landau,Gilbert,Slonczewski96,Everschor2011,Zhou19,Tomasello14,Taniguchi,Berkov2008,Menezes1,Menezes2,Mulkers17,Yuan19}
\begin{equation}
\frac{d\bm{m}}{dt} = -\gamma \bm{m} \times B_{\rm eff} + \alpha \bm{m} \times \frac{d\bm{m}}{dt} - \beta \bm{m}\times [\bm{m} \times \bm{j}_{\rm p}]
\label{LLGS}
\end{equation}
where $\bm{m}$ is the magnetization unit vector, $\gamma >0$ is the gyromagnetic ratio, $\alpha$ is the Gilbert damping constant defining dissipation term, the constant $\beta$ parametrizes the spin-transfer-torque, and $\bm{j}_{\rm p}$ is the applied spin-polarized current. Following the approach of Thiele, by neglecting any deformation \cite{Knoester} of the spin-structure of a skyrmion during its motion, the LLGS equation can be mapped onto the so called Thiele equation \cite{Thiele73,Thiele74,Liu,Kim19,Zhou19,Tomasello14,Menezes1,Menezes2,Iwasaki14,Yuan19}:
\begin{equation}
{\cal G}_{ij} v_j -\alpha {\cal D}_{ij} v_j = F_i
\label{Thiele}
\end{equation} 
where $v_j$ is the velocity of the spin-texture. The gyroscopic and dissipative force tensors and driving force vector are respectively given by
\begin{eqnarray}
{\cal G}_{ij} &=& \frac{1}{4\pi}\int d^2\bm{r}\,  \left( \frac{\partial \bm{m}}{\partial x_i} \times \frac{\partial \bm{m}}{\partial x_j} \right) \bm{\cdot}\bm{m} \\
{\cal D}_{ij} &=& \frac{1}{4\pi}\int d^2\bm{r}\,  \left( \frac{\partial \bm{m}}{\partial x_i} \bm{\cdot} \frac{\partial \bm{m}}{\partial x_j} \right) \\
F_i &=& \frac{\beta}{4\pi}\int d^2\bm{r}\, \left( \frac{\partial \bm{m}}{\partial x_i} \times  \bm{m} \right) \bm{\cdot} \bm{j}_{\rm p}
\label{force}
\end{eqnarray}

{\it Skyrmion Hall angle}:
Expressing $\bm{m}(\bm{r})$ in spherical variables $(\Theta (\bm{r}), \Phi (\bm{r}))$ of unit radius and solving the corresponding Euler equations, we recently have reported that a very accurate skyrmion solution \cite{BM} is
\begin{equation}
\Theta (r) = 4 \arctan \left( \exp \left[-\delta \frac{r}{r_s}\right] \right)
\label{Sk_sol}
\end{equation}
and $\Phi (\phi) = \phi $ for N\'eel type skyrmioms, where $r_s = \frac{D}{H}$ defined as the ratio the strength of Dzyaloshinskii-Moriya interaction $D$ and applied magnetic field $H$ fixes the length scale of a skyrmion and the dimensionless parameter $\delta$ depends on the ratio of magnetic anisotropy $A$ and $H$. 

The skyrmion solution (\ref{Sk_sol}) yields 
\begin{equation}
{\cal G} = \left( \begin{array}{rr} 0 & -  Q \\  Q & 0 \end{array} \right) \, , \,\,
{\cal D} = \left( \begin{array}{rr} S & 0 \\ 0 & S \end{array} \right) \, ,
\label{Parameter}  
\end{equation}
and $\bm{F} = {\cal F} \left(j_{{\rm py}},\, -j_{{\rm px}} \right)$.  We find (using Mathematica\cite{Mathematica})
\begin{equation}
{\cal F}=-\frac{1}{4} \int_0^\infty dr\left[ r \frac{d\Theta}{dr} + \frac{\sin 2\Theta}{2} \right]
 =(G-\frac{1}{6})\frac{\beta r_s}{\delta}
\end{equation}
 with Catalan's constant $G=\sum_{l=0}^\infty \frac{(-1)^l}{(2l+1)^2} \approx 0.92$, topological charge $Q = +1$ for skyrmions, 
\begin{eqnarray}
	S &=& \frac{1}{4} \int_0^\infty dr\left[ r\left( \frac{d\Theta}{dr}\right)^2 + \frac{\sin^2\Theta}{r} \right] \label{expression_s} \\
	&=&  \int_0^\infty dx \left[ \frac{x e^{x}}{(e^{x}+1)^2} + 4 \frac{e^{x}(e^{x}-1)^2}{x(e^{x}+1)^4} \right]
	\label{expression_s2}
\end{eqnarray}
where dimensionless $x= 2\delta (r/r_s)$. Integrating by parts for the second integral in Eq.~(\ref{expression_s2}), we find
\begin{eqnarray}
	 S &=&   \ln(2) + 8\int_0^\infty \frac{\tanh^3 (x/2) {\rm sech}^2(x/2)}{x(e^x-1)} dx \nonumber \\
	&&-4 \int_0^\infty \frac{\tanh^4(x/2)}{x^2(e^x-1)} dx  \\
	&=& \left(\frac{41}{45}\right)\ln (2) - 4 \zeta'(-1) -40 \zeta'(-3)
	\label{analytical_s}
\end{eqnarray}
obtained using Mathematica\cite{Mathematica} with $\zeta'(x)$ being the derivative of Riemann zeta function $\zeta (x) = \sum_{n=1}^\infty \frac{1}{n^x}$.  We here note that the constant  $S \approx 1.08$ is independent of the length-scale $r_s$. 

The Euler equation for skyrmions with the scaled variable $\rho = r/r_s$ is given by \cite{BM}
\begin{eqnarray}
	&&\frac{d^2\Theta}{d\rho^2} + \frac{1}{\rho}\frac{d\Theta}{d\rho}- \frac{\sin (2\Theta) }{2\rho^2}   \nonumber \\
	&& =  \frac{H_0}{H} \left( -\frac{2}{\rho}\sin^2\Theta 
	+\sin\Theta  - \frac{A}{H} \sin(2\Theta) \right)    
	\label{Euler}
\end{eqnarray}
where $H_0 =A_0 = \frac{D^2}{J}$, with $J$, $D$, $H$, and $A$ respectively being the strengths of exchange energy, Dzyaloshinski-Moriya interaction, magnetic field, and easy-axis or easy-plane magnetic anisotropy in a continuum model \cite{BM,Leonov16}.  Equation (\ref{Sk_sol}) is an approximate closed-form solution of Eq.(\ref{Euler}) for almost entire region of parameter space of $\frac{H}{H_0}$ and $\frac{A}{A_0}$ for which skyrmions describe \cite{BM} the ground state. Using numerical solution of Eq.(\ref{Euler}), we here determine the value of $ S$ for many different set of these parameters within the skyrmion phase (Fig.~2a) in the phase diagram. The corresponding values of $S$ are shown in Fig.~2b. Apart from the region of relatively small values of $\frac{H}{H_0}$, the value of $S$ satisfactorily  agrees with the analytical result (\ref{analytical_s}). However, any deviation from the analytical value of $S$ dose not contradict that $S$ is length-scale independent even for smaller $\frac{H}{H_0}$, because it is the consequence of Euler equation (\ref{Euler}) and the expression (\ref{expression_s}) of $S$. The change in $S$ occurs due to change in profile/shape of the skyrmion; $S$ remains constant for small to large skyrmions if their shapes are same. 
Equation (\ref{expression_s}) suggests that if $\Theta (r)$ depends on single length scale which is precisely the case of the Euler equation (\ref{Euler}), $ S$ is length-scale independent. We note that previous estimations
\cite{Zhou19,Jiang17,Juge19,Litzius2017,Liu19,Yuan19} of $ S$ are made using $\Theta (r)$ consisting of two length scales leading to length-scales dependent $ S$.
The fact that $ S$ is length-scale independent for a solution of $\Theta (r)$ depending on single length scale   has a far reaching consequence in the length-scale-independent $\theta_{\rm sk}$ as we show below.

\begin{figure}
\includegraphics[scale=0.95]{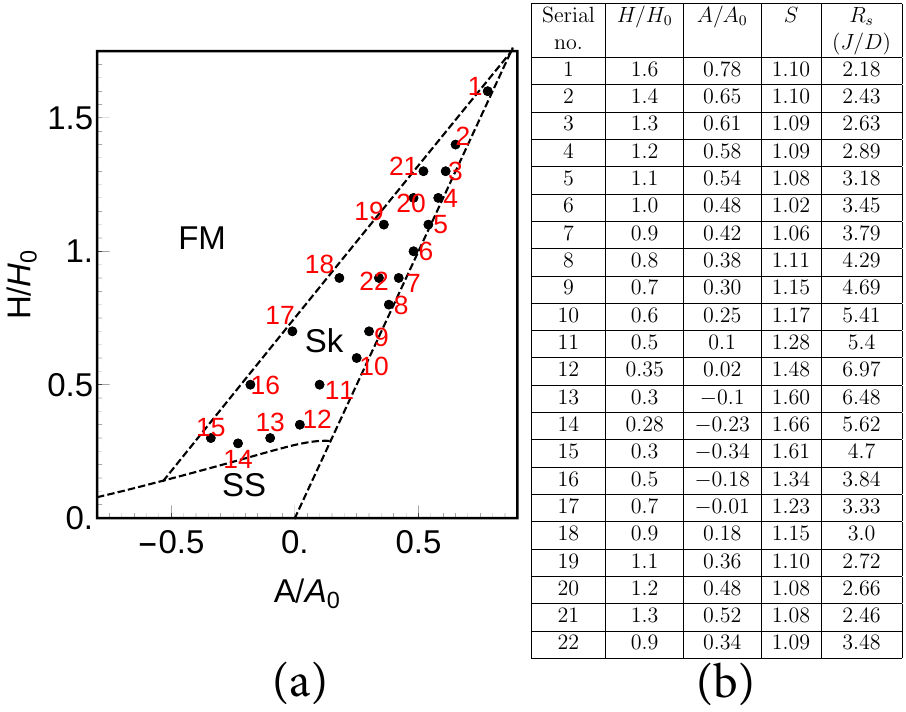}
\caption{(a) In the skyrmion (Sk) phase of the phase diagram \cite{BM}, some points are assigned at which $ S$ is calculated using numerical solution of the appropriately scaled Euler equation (\ref{Euler}). Here FM and SS respectively represent ferromagnetic and spin-spiral phases. (b) The values of $\frac{H}{H_0}$, $\frac{A}{A_0}$ and the corresponding calculated $S$ are listed for the marked points in (a). The tabulated values of the radii of skyrmions, $R_s$, have been calculated using an expression provided in Ref.~\onlinecite{BM}. }
\end{figure}

The matrix form of Thiele's equation for N\'eel type skyrmions can thus be written as
\begin{equation}
\left[ \begin{array}{rr} -\alpha S & - Q \\ Q & -\alpha S \end{array} \right] \left( \begin{array}{c} v_x \\ v_y \end{array} \right) = {\cal F}\left( \begin{array}{r} j_{{\rm py}} \\ -j_{{\rm px}}  \end{array} \right) 
\end{equation}
whose solution is found to be
\begin{equation}
v_x = -{\cal F}\left( \frac{\alpha S j_{{\rm py}} + Q j_{{\rm px}}}{Q^2 + \alpha^2 S^2} \right)\, ,\,
v_y = {\cal F}\left( \frac{\alpha S j_{{\rm px}} - Q j_{{\rm py}}}{Q^2 + \alpha^2 S^2} \right)\, .
\label{velocity}
\end{equation}
As ${\cal F}$ depends on the length scale $r_s$ of the skyrmions, their velocities (\ref{velocity}) are length scale dependent, as found in the experiments \cite{Litzius2017,Legrand2017,Ding}.

For the skyrmion-Hall effect, we set $j_{{\rm py}} =0$ and the tangent of the corresponding Hall angle $\theta_{\rm sk}$ is determined by the ratio $\frac{v_y}{v_x}$. We thus find
\begin{equation}
\tan \theta_{\rm sk} = -\frac{\alpha S}{Q} \,.
\label{Hall_amgle}
\end{equation}  
Therefore, the skyrmion Hall angle is length-scale $r_s$ independent.
The radius of a skyrmion, $R_s$, will depend on sole length scale $r_s$ as $R_s = r_s w$, where the dimensionless factor $w$ may depend on the dimensionless parameters of the system. If we absorb $w$ in the redefined length scale as $r'_s  = w r_s$, $\theta_{\rm sk}$ becomes independent of $r'_s$ too and thus the skyrmion Hall angle becomes insensitive to the radius of a skyrmion, as reported by Zeissler et al \cite{Zeissler20}. On the other hand, $\theta_{\rm sk}$  may change due to change in $S$ when the profile/shape of the skyrmion alters if one changes  magnetic field from low to high values or vice versa. However, for a particular shape of the skyrmion, $\theta_{\rm sk}$ is independent of its size.
Since $Q= +1$ and $S \approx 1.08$ is a numerical constant, we readily estimate $\alpha \approx 0.11$--$0.18$ for the diameter-independent measured skyrmion Hall angle $9^{\rm o}\pm 2^{\rm o}$ for N\'eel type skyrmions by Zeissler et al.\cite{Zeissler20}. As reported in the experiments, $\theta_{\rm sk}$ is also found to be input current density independent.

By setting $j_{{\rm py}} = \frac{\alpha S}{ Q} j_{{\rm px}}$, we find $v_y = 0$ and $v_x = -\frac{{\cal F}}{Q}j_{{\rm px}}$. In other words, the skyrmion Hall angle vanishes in such a condition and thus the skyrmions will not at all move along transverse direction. The mechanisms for restoring the motion of skyrmions along a particular direction is long-sought for technological applications. We here propose that application of an appropriate current along transverse direction will balance (Fig~1) the motion of skyrmions along transverse direction. If $j_{{\rm py}} > \frac{\alpha S}{ Q} j_{{\rm px}}$, $\theta_{\rm sk}$ will reverse its sign, i.e., the skyrmions' transverse motion will be in the opposite direction. 
In this paper, the current densities are assumed to be constant all throughout the system and hence the skyrmions are expected to have uniform motion. 
In the real systems with the proposed setup (Fig.~1), the current densities may become nonuniform towards the edges. However, so long as the degree of inhomogeneities is not large such that linear dependency of current densities in Thiele equation remains unchanged, the local Hall angle will depend on the ratio of local components of the current densities and remains length-scale independent. Although the motion of skyrmions will not be uniform, a desired small Hall angle condition can still be achieved due to this.

{\it Meron Hall angle}:
Another topological spin structure has been observed as merons \cite{Phatak12,Zhang20} with $Q_{\rm mr}= \frac{1}{2}$. Merons are formed in chiral magnets with moderate easy-plane magnetic anisotropy \cite{BM,BM19} in the absence of magnetic field. While spin-orientation  far away from the center of a skyrmion is in a plane but opposite in direction along a line, the center of a meron will have spin-orientation either up or down to the plane in consideration. To match with the underlying planar ferromagnetic spin-orientation, these merons must form a lattice \cite{BM,Liu} with consecutive merons ought to have opposite spin-orientation at their respective centers. According to Ref.\cite{BM}, N\'eel type meron solutions are given by $\Phi(\phi) = \phi$ and
\begin{equation}
	\Theta_{\rm mr} (r) = 2 \arctan \left( \exp \left[-\delta' \frac{r}{r_a}\right] \right) \pm \frac{\pi}{2}
	\label{meron}
\end{equation} 
with length-scale $r_a = D/A$ and $\delta' =\frac{2\ln(2)}{1+2G} \approx 0.49 $. Here $\pm$ signs refer to spin down and up at the center of the merons which we refer below as down-meron and up-meron  respectively. Although a single meron solution (\ref{meron}) suggests that far away from its center $\Theta_{\rm mr} (r\to \infty) = \pm \frac{\pi}{2}$, there will not be large regime of $r$ for which $\Theta_{\rm mr} (r) \approx \pm \frac{\pi}{2}$ because of the interactions between up and down-merons as they form lattice.

Similar to the calculation for skyrmions above, we find 
${\cal F}_{\rm mr} =\frac{(1+2G)\beta r_a}{4\delta'}$ for a meron.
The dissipation constant, $S_{\rm mr}$,  for a meron is given by
\begin{eqnarray}
	 S_{\rm mr} &=& \frac{1}{4} \int_0^\infty dr\left[ r\left( \frac{d\Theta_{\rm mr}}{dr}\right)^2 + \frac{\sin^2\Theta_{\rm mr}}{r} \right] \nonumber \\
	&=&   \frac{1}{4}\int_0^\infty dx \left[ \frac{x e^{x}}{(e^{x}+1)^2} +  \frac{(e^{x}-1)^2}{x(e^{x}+1)^2} \right]
	\label{meron_s}
\end{eqnarray}
with $x= \frac{ 2 \delta' r}{r_a}$. The  second term in Eq.(\ref{meron_s}) is logarithmically divergent for a single meron because for  $r \gg r_a$, $\Theta_{\rm mr} \simeq \pm \frac{\pi}{2}$. However, as the merons form a lattice,  $\Theta_{\rm mr} = \pm \frac{\pi}{2}$ does not occur for a large range of $r$ and thus we can apply a suitable cut-off in the corresponding integral in Eq.(\ref{meron_s}). Considering the cut-off radius $R= 15 r_a$ that corresponds to $\Theta_{\rm mr} (R) \approx \pi \frac{700}{1399}$ which is nearly equal to its asymptotic value $\frac{\pi}{2}$, we find $S_{\rm mr} \approx 0.72$. Actual value of $S_{\rm mr}$ in the meron lattice is expected to be smaller than the value estimated here as the value of $R$ considered here is reasonably large. 
While actual estimation of  $S_{\rm mr}$ for a meron lattice will depend on the interaction between merons, it is indeed beyond the scope of the present paper as our primary aim is to show the length-scale independent meron Hall angle. For a meron lattice, $\Theta_{\rm mr} (r) \to \sum_j \Theta_{\rm mr} (\vert {\rm \bf r}-{\rm \bf X}_j\vert)$ where ${\rm \bf X}_j$ is the position of a meron in a lattice and thus apparent singularity in Eq.(\ref{meron_s}) is removed for a lattice maintaining scale independent $S_{\rm mr}$.


The meron-Hall angles for up-meron and down-meron respectively are then found to be
\begin{equation}
	\tan \theta_{\rm mr}^\uparrow = - \tan \theta_{\rm mr}^\downarrow \approx \frac{\alpha S_{\rm mr}}{Q_{\rm mr}} \, .
\end{equation}  
As the signs of ${\cal G}$ matrices are opposite for up and down merons, the corresponding Hall angles are opposite in signs.
Therefore they will move in opposite directions, as
found in a recent experiment \cite{Hirata2019}. As a result, a net spin-Hall effect will be present in meron lattice subjected to the application of in-plane current. Like skyrmion-Hall angle, meron-Hall angle is found to be the length scale and input current density independent as is observed in an experiment \cite{Zhang20}.  

{\it Conclusion}: We  have shown here that the skyrmion and meron Hall angles are independent of the input current density and the corresponding length scales provided the shape of these topological structures remains same, regardless of their sizes. The Hall angle may change for one shape to another, but the degree of change depends on the value of dissipative parameter $S$.  
If there occurs a large change in the shape of the skyrmion due to the wide variation of the magnetic field so that $S$ changes quite a bit, yet the Hall angle remains same, the pinning effect may find further role  \cite{Muller2015,Reichhardt2015,Diaz2017,Reichhardt2015b,Stosic17,Hanneken,Bhattacharya}. 

Our results are also valid for asymmetric skyrmions \cite{Yu2018} as well as merons. Our proposal for realizing zero skyrmion Hall angle will pave the way in device  applications  based on skyrmions. Similar to N\'eel type skyrmions, Hall angle for  Bloch type skyrmions will also be independent of length scale and input current density.

\end{document}